\documentclass[aps,pra,reprint,superscriptaddress,nofootinbib]{revtex4-2}
\usepackage[colorlinks=true,linkcolor=blue,citecolor=blue,urlcolor=blue]{hyperref}
\usepackage{graphicx}
\usepackage{amsmath}
\usepackage{amssymb}
\usepackage{mathrsfs}
\usepackage{xcolor}
\usepackage{mathtools}
\usepackage{orcidlink}
\usepackage{braket,bm}
\usepackage{longtable}
\usepackage{verbatim}
\usepackage{gensymb}

\usepackage{multirow}
\usepackage{makecell}

\newcommand{\rmel}[3]{\langle #1 \Vert #2 \Vert #3 \rangle}

\begin{document}


\title{Simulated Laser Cooling and Magneto-Optical Trapping of Group IV Atoms}

\author{Geoffrey Zheng\:\orcidlink{0000-0002-9312-0102}}
\thanks{These authors contributed equally to this work.}
\affiliation{James Franck Institute and Department of Physics, University of Chicago, Chicago, Illinois 60637, USA}

\author{Jianwei Wang\:\orcidlink{0009-0001-3039-2602}}
\thanks{These authors contributed equally to this work.}
\affiliation{James Franck Institute and Department of Physics, University of Chicago, Chicago, Illinois 60637, USA}
\affiliation{William H. Miller III Department of Physics and Astronomy, Johns Hopkins University, Baltimore, Maryland 21218, USA}

\author{Mohit Verma\:\orcidlink{0000-0001-7548-3930}}
\affiliation{James Franck Institute and Department of Physics, University of Chicago, Chicago, Illinois 60637, USA}

\author{Qian Wang\:\orcidlink{0000-0003-2853-8534}}
\altaffiliation{Present address: PsiQuantum, Inc., Milpitas, CA 95035, USA}
\affiliation{James Franck Institute and Department of Physics, University of Chicago, Chicago, Illinois 60637, USA}

\author{Thomas K. Langin\:\orcidlink{0000-0001-9234-3825}}
\altaffiliation{Present address: IonQ, Inc., Bothell, WA 98021, USA}
\affiliation{James Franck Institute and Department of Physics, University of Chicago, Chicago, Illinois 60637, USA}

\author{David DeMille\:\orcidlink{0000-0001-7139-4121}}
\email[Correspondence should be addressed to: ]{david.demille@jhu.edu; geoffreyz@uchicago.edu; jwang695@jh.edu}
\affiliation{James Franck Institute and Department of Physics, University of Chicago, Chicago, Illinois 60637, USA}
\affiliation{William H. Miller III Department of Physics and Astronomy, Johns Hopkins University, Baltimore, Maryland 21218, USA}

\date{\today}

\begin{abstract}
We present a scheme for laser cooling and magneto-optical trapping of the Group IV (a.k.a. Group 14 or tetrel) atoms silicon (Si), germanium (Ge), tin (Sn), and lead (Pb). These elements each possess a strong Type-II transition ($J \rightarrow J' = J-1$) between the metastable $s^2p^2 \,^3\!P_1$ state and the excited $s^2ps'\, ^3\!P_0^\circ$ state at an accessible laser wavelength, making them amenable to laser cooling and trapping. We focus on the application of this scheme to Sn, which has several features that make it attractive for precision measurement applications. We perform numerical simulations of atomic beam slowing, capture into a magneto-optical trap (MOT), and subsequent sub-Doppler cooling and compression in a blue-detuned MOT of Sn atoms. We also discuss a realistic experimental setup for realizing a high phase-space density sample of Sn atoms.
\end{abstract}

\maketitle

\section{Introduction}

Laser cooling and magneto-optical trapping of neutral atoms \cite{HanschLaserCoolingProposal, PhillipsZeemanSlower, ChuOpticalMolasses, ChuMOT} has revolutionized modern atomic, molecular, and optical (AMO) physics. Magneto-optical traps (MOTs) are the workhorse of ultracold atomic physics, enabling realization of atomic Bose-Einstein condensates and Fermi degenerate gases \cite{CornellBEC, KetterleBEC, JinFermiDegenerateGas}, loading of optical tweezer arrays \cite{BernienTweezerArray, EbadiAtomArray1}, and precision quantum metrology \cite{HidetoshiKatoriGravRedshift, JunYeGravRedshift, MuellerAtomInterferometer, VuleticEntangledAtomicClock}. Only a limited subset of elements on the periodic table have been laser cooled because of the requirement for repeated photon cycling when exciting the atom from a long-lived (ground or metastable) state $\ket{g}$ to a short-lived excited state $\ket{e}$. Laser cooling of atoms is nearly always realized using a Type-I transition, where the total angular momentum $F$ ($F'$) of $\ket{g}$ ($\ket{e}$) have the relation $F'=F+1$. Such transitions with accessible laser wavelengths are found in atoms of alkali and alkaline earth metals, metastable noble gases, plus certain rare earths~\cite{JapanYbMOT, ErbiumMOT, DyMOTLev, HoMOTSaffman, RussiaThuliumMOT, JapanEuropiumMOT}, transition metals~\cite{CrMOT, CdMOT, GermanyAgMOT, JapanHgMOT,StamperKurnTransitionMetalMOTs, StamperKurnTitaniumMOT, NorrgardMoMOT}, and Group III (triel) elements~\cite{NicholsonIndiumMOT}.

Laser cooling and magneto-optical trapping have also been applied to molecules, unlocking a new class of quantum matter to explore. Molecular structure dictates that optical cycling must operate on a Type-II transition ($F'=F-1$ or $F' = F$) for rotational closure. To date, several molecular species have been laser cooled and trapped in a MOT~\cite{DeMilleSrFMOT,TarbuttCaFMOT, YeYOMOT, DoyleCaOHMOT, DoyleSrOHMOT, YanBaFMOT, TruppeAlFMOT}. Among these molecules, CaF, SrF, YO, CaOH, and SrOH have been further cooled to ultracold temperatures ($\sim\! 5 - 100\,\mu\text{K}$) and loaded into conservative optical dipole traps (ODTs) \cite{DoyleCaFODT, DeMilleSrFODT, YeYOCollisions, CaOHODT, SrOHODTandCBMOT}. These developments indicate that Type-II transitions are well-suited for use in various quantum science applications.

Adding a new class of elements to the list of laser-coolable species would constitute another leap in the quest to control and manipulate quantum matter. In this work, we present numerical simulations to verify a scheme for laser cooling and trapping of nearly all\footnote{The vacuum UV transition wavelength for carbon (C) cannot currently be produced with sufficient continuous-wave power for laser cooling and trapping (see Table \ref{tab:Group IV cycling transitions}), so we do not consider it here.} Group IV elements --- silicon (Si), germanium (Ge), tin (Sn), and lead (Pb) --- using a Type-II transition. Among these elements, previous attempts have been made to study pathways towards producing ultracold gases of Si atoms \cite{SiLaserCoolingJapan, SALeeSiLaserCooling, RonaldSiLaserCoolingThesis, USNASiSpectroscopy}, but a Si MOT has not been realized to date. Our work should provide a general path forward for laser cooling and trapping of Group IV elements for applications in quantum science and precision measurement.

\section{Laser Cooling and Trapping Scheme}

\subsection{Structure of Group IV Atoms}

\begin{figure}
    \includegraphics[width=0.8\linewidth]{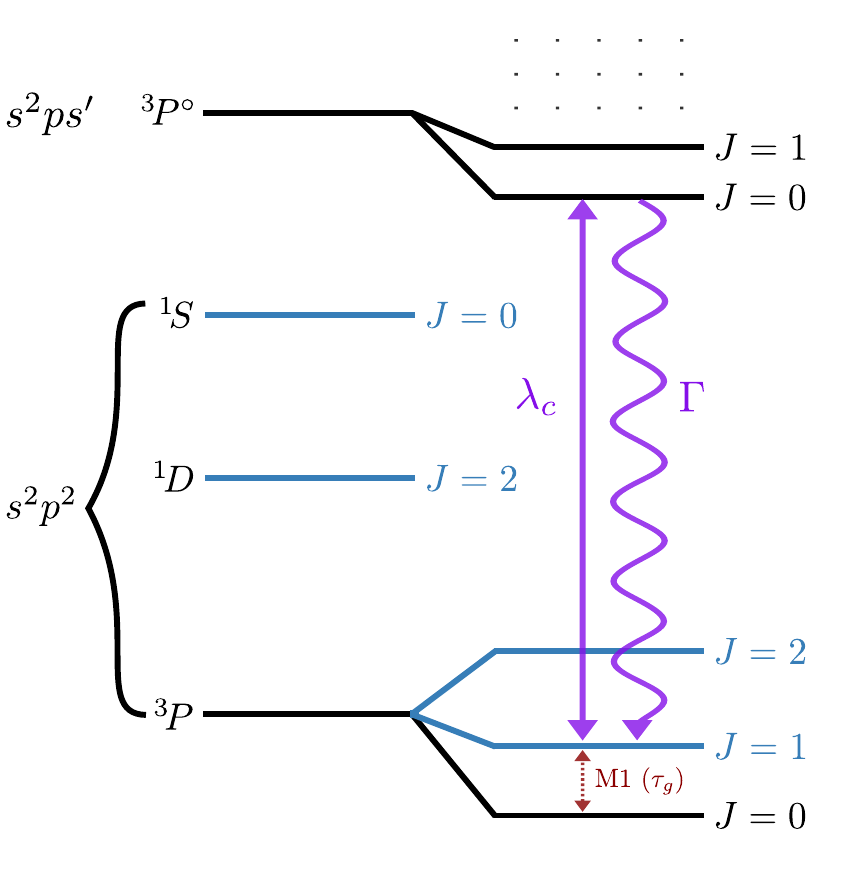}
    \caption{Lowest-lying energy levels of Group IV elements relevant for laser cooling/trapping and precision measurement. Light blue lines indicate metastable states in the ground $s^2p^2$ configuration. We model laser cooling and trapping of these elements using a single laser on the 
    $\protect\ket{s^2p^2\, {^3}\!P_1} \rightarrow \protect\ket{s^2ps'\,{^3}\!P_0^\circ}$ closed optical cycling transition (wavelength $\lambda_c$, natural linewidth $\Gamma$). The metastable $\ket{s^2p^2\, {^3}\!P_1}$ state (lifetime $\tau_g$) can only decay to $\ket{s^2p^2\, {^3}\!P_0}$, via an M1 transition. Dotted lines indicate higher lying states not relevant to the cooling and trapping scheme.}
    \label{fig: tin level diagram}
\end{figure}

In their ground state, all Group IV elements have four valence electrons in a $s^2p^2$ atomic orbital configuration. This leads to five even parity states in the ground-state manifold, with terms $^3\!P_{0,1,2}, {^1\!D}_2$, and $^1\!S_0$ in order of energy from lowest to highest. The four excited states in this manifold cannot decay via electric dipole (E1) radiation and hence are all long-lived. The lowest-lying odd parity excited state has configuration $s^2ps'$ and term $^3\!P_0^\circ$.\footnote{C and Si also have a $sp^3$ $^5\!S_2^\circ$ state below the $s^2ps'$ $^3\!P^\circ$ states and above the $s^2p^2$ $^{1}\!S_0$ state, but this does not affect our discussion.}  
A level diagram of this structure is shown in Fig.~\ref{fig: tin level diagram}.

We propose using the metastable $s^2p^2\, {^3}\!P_1$ to excited $s^2ps'\,{^3}\!P_0^\circ$ transition in the Group IV elements as a pure Type-II optical cycling transition that is amenable to laser cooling and trapping. This transition is dipole-allowed with large natural linewidth. Moreover, it is completely closed by selection rules, meaning that no repumper laser is necessary.\footnote{In Pb, the metastable state lifetime is short enough that a $^3\!P_0 \rightarrow {^3}\!P_1$ repumper laser ($\lambda = 1279$ nm) may be useful.}  Table \ref{tab:Group IV cycling transitions} lists the relevant properties of this transition across all Group IV elements, as well as the metastable state lifetime. In the case of Si, Ge, Sn, and Pb, the transition wavelength can be produced with sufficient power using current UV laser technology. 

\begin{table}
\caption{Optical transition wavelength $\lambda_c$, natural linewidth $\Gamma$, and saturation intensity $I_\text{sat}$ of the $s^2p^2\,{^3}\!P_1 \rightarrow s^2ps'\,{^3}\!P_0^\circ$ transition proposed for use in laser cooling and trapping of Group IV elements \cite{NIST_ASD}. The metastable $s^2p^2\,{^3}\!P_1$ state lifetime $\tau_g$ is also listed~\cite{NIST_ASD}. Details on calculating $\tau_g$ are given in Appendix \ref{sec:AppxMetastableLifetime}.}
\label{tab:Group IV cycling transitions}
\begin{ruledtabular}
\begin{tabular}{c c c c c}
Element & $\lambda_c$ (nm) & $\Gamma/2\pi$ (MHz) & $I_\text{sat}$ (mW/cm$^2$) & $\tau_g$ (s) \\
\hline
C  & 166 & $55$ & 1600 & $1\times 10^7$ \\ 
Si & 252 & $35$ & 290 & $1\times 10^5\!$ \\
Ge & 271 & $45$ & 300 & $3 \times 10^2$ \\
Sn & 303 & $32$ & 150 & 11 \\ 
Pb & 368 & $22$ & 60 & 0.14 \\
\end{tabular}
\end{ruledtabular}
\end{table} 

\subsection{Applications of Laser-Cooled Group IV Atoms}
Ultracold atomic gases of Group IV elements have potential applications ranging from improved quantum sensing and novel quantum computing platforms to precision tests of fundamental physics and searches for physics beyond the Standard Model (BSM). Initial efforts to produce an ultracold atomic gas of Si atoms focused on envisioned applications to solid-state-based nuclear spin quantum computing \cite{SiKaneQuantumComputer, SiQuantumComputer} and quantum sensing \cite{SiQuantumSensing}. Resonant photoionization of ultracold radioactive $^{31}\text{Si}$ atoms in a MOT, followed by ion-guiding optics, would enable nm-scale precision implantation of $^{31}\text{Si}^+$ ions in a Si substrate~\cite{RonaldSiLaserCoolingThesis, McClellandIonImplantationMOT}. Subsequent $\beta$-decay of these nuclei ($\tau \sim\! 3$ hrs) would create an array of $^{31}\text{P}^+$ ions doped in the Si substrate, realizing the building blocks of the Kane quantum computing architecture~\cite{SiKaneQuantumComputer}. 

In addition to supporting a Type-II cycling transition, all Group IV elements possess various clock transitions that can address all four linearly independent energy differences between the long-lived $s^2p^2$ states. Details on these transitions, which can only be of magnetic dipole (M1) or electric quadrupole (E2) type, are listed in Table~\ref{tab:Group IV clock transitions} for the case of Sn.

\begin{table}
\caption{List of narrow linewidth clock transitions between $s^2p^2$ states in Group IV elements. Asterisk (*) denotes nominally spin-forbidden transitions. For the specific case of Sn, we list the wavelength $\lambda$~\cite{NIST_ASD}, calculated partial linewidth $\Gamma_\text{p}$~\cite{BiemontAstrophysForbiddenTransition}, and total linewidth $\Gamma_\text{tot}$.}
\label{tab:Group IV clock transitions}
\begin{ruledtabular}
\begin{tabular}{c c c|| c c c}
  \multicolumn{3}{c||}{} & \multicolumn{3}{c}{\bf Atomic Sn} \\
  \cline{4-6}
  State & Transition & Type & $\lambda$ (nm) & $\Gamma_\text{p}/2\pi$ (Hz) & $\Gamma_\text{tot}/2\pi$ (Hz) \\
  \hline
  $^3\!P_1$ 
    & $\leftrightarrow\,^3\!P_0$ 
    & M1        
    & 5911 
    & $ 0.01$ 
    & $ 0.01$ \\
  \hline
  \multirow{2}{*}{$^3\!P_2$} 
    & $\leftrightarrow\,^3\!P_0$ 
    & E2        
    & 2917 
    & $ 10^{-4}$ 
    & \multirow{2}{*}{$ 0.01$} \\
  & $\leftrightarrow\,^3\!P_1$ 
    & M1/E2     
    & 5761 
    & $ 0.01$ 
    &                             \\
  \hline
  \multirow{3}{*}{$^1\!D_2$} 
    & $\leftrightarrow\,^3\!P_0$ 
    & \:\! E2*  
    & 1161 
    & $ 10^{-5}$ 
    & \multirow{3}{*}{$ 0.16$} \\
  & $\leftrightarrow\,^3\!P_1$ 
    & \:\! M1/E2* 
    & 1445 
    & $ 0.07$ 
    &                             \\
  & $\leftrightarrow\,^3\!P_2$ 
    & \:\! M1/E2*
    & 1929 
    & $ 0.08$ 
    &                             \\
  \hline
  \multirow{3}{*}{$^1\!S_0$} 
    & $\leftrightarrow\,^3\!P_1$ 
    & \:\! M1*  
    & 646  
    & $ 1.13$ 
    & \multirow{3}{*}{$ 1.4$} \\
  & $\leftrightarrow\,^3\!P_2$ 
    & \:\! E2*    
    & 728  
    & $ 0.11$ 
    &                             \\
  & $\leftrightarrow\,^1\!D_2$ 
    & E2        
    & 1170 
    & $ 0.18$ 
    &                             \\
\end{tabular}
\end{ruledtabular}
\end{table}

The plethora of metastable states in the ground $s^2p^2$ configuration provides several accessible clock transitions that can be utilized for a variety of precision measurement experiments. Sn is particularly attractive for such applications because it possesses a chain of seven stable $I=0$ isotopes (the longest found in nature). The combination of many clock transitions and many $I=0$ isotopes makes laser-cooled and trapped Sn atoms a promising platform for probing non-linearity in King plot isotope shift measurements \cite{BerengutIsotopeShiftSpectroscopy, FuchsIsotopeShiftSearch}. Detection of non-linearity could indicate BSM physics manifesting as a new, light boson that interacts with electrons and nucleons~\cite{BudkerIsotopeShifts}. Other charge states of Sn, such as $\text{Sn}^{2+}$ and $\text{Sn}^{7+}\text{-}\text{Sn}^{10+}$ ions, have been recently explored and also possess suitable clock transitions for such measurements~\cite{leibrandt2024prospects, torretti2017opticalHighChargedSn, YuForbiddenHighlyChargedIons, LyuUltrastableXUVChargedIons}.

The properties of Group IV elements also make them attractive for the measurement of atomic parity violation (APV). Measurements of ratios of APV amplitudes in a long, stable isotope chain make it possible to cancel uncertainty from electronic structure factors that are difficult to compute from \textit{ab initio} theory~\cite{FlambaumAPVRatios, WiemanAPVRatios, FortsonAPVRatios}. Clock transitions greatly improve energy resolution, and hence can provide a platform for more precise measurements of APV than in prior experiments~\cite{IonAPVFortson}. Together, these features open a path to more sensitive probes of the nuclear weak charge~\cite{SafronovaDeMilleFundamentalPhysicsReview, AntypasBudkerYbIsotopeAPVChain}. Alternatively, assumption of Standard Model values for weak interaction parameters enables measurements of APV along a chain of $I=0$ isotopes to precisely probe the ``neutron skin" effect ~\cite{PhysRevC1992,NeutronSkinPhysRevCFlambaum2009,PhysRevC2019Flambaum}. Combined with other nuclear physics experiments, such measurements may provide critical insight into the equation of state of neutron-rich matter and establish a direct link between nuclei and neutron stars \cite{NeutronSkinReview}. Isotopes with nonzero nuclear spin can also probe nuclear spin-dependent parity violation (NSD-PV), including contributions from the nuclear anapole moment where there is tension between nuclear theory and experimental measurements \cite{HaxtonWiemanAnapoleReview}. 

Finally, laser cooling of Group IV atoms introduces the opportunity to assemble ultracold, diatomic molecules built from Group IV and coinage metal atoms (Cu, Ag, Au), where laser cooling has been demonstrated (for Ag) ~\cite{GermanyAgMOT, ChicagoAgMOT}. This unique class of molecule was recently proposed as a promising candidate for next-generation precision measurements of the electron electric dipole moment (eEDM)~\cite{BenPbAgEDM}.

\subsection{Scheme for Laser Cooling and Trapping with the $^3\!P_1 \rightarrow {^3\!P}_0^\circ$ Transition}\label{sec:LaserCoolingSchemeSpinZero}
All Group IV elements have multiple spin-zero ($I=0$) even isotopes and several spinful odd isotopes; in this paper we focus only on the even isotopes, where laser cooling and trapping is simplified by their lack of hyperfine structure. The Type-II optical cycling transition $\ket{^3\!P, J=1} \rightarrow \ket{^3\!P^\circ, J'=0}$ has three ground-state Zeeman sublevels and one excited-state Zeeman sublevel. This is a conceptually clean system to study numerically and stands in marked contrast to Type-II optical cycling transitions used in molecular laser cooling, where hyperfine structure and vibrational state leakage lead to much more complexity.

We note, in passing, that a more conventional-seeming optical cycling transition is also present: $s^2p^2\,^3\!P_2 \leftrightarrow sp^3\,{^3}\!D_3^\circ$ (for Si) or $s^2pd\,{^3}\!D_3^\circ$ (for Ge and Sn), i.e. the excited state is the lowest-lying odd-parity triplet state with $J=3$. This was used to attempt laser cooling of Si ($\lambda = 222\,\text{nm}$) \cite{RonaldSiLaserCoolingThesis}. This cycling transition is leaky, with an estimated branching ratio $^3\!D_3^\circ \rightarrow s^2p^2 {^1}\!D_2$ of $\sim\! 0.02\% - 50\%$ depending on the element \cite{NIST_ASD}, necessitating a repumper laser to close the cycling transition. However, including this repumper creates a $\Lambda$-system, effectively making a Type-II cycling transition with two UV lasers on two different transitions. Our approach is simpler and should lead to more effective laser cooling and trapping of Group IV elements. 

The presence of dark Zeeman sublevels in the ground state necessitates a mechanism to destabilize such states during optical cycling, lest population accumulate there and cause cooling and trapping to cease~\cite{BerkelandBoshierDarkStateRemix}. It also leads to reduced photon scattering rate and optical force compared to conventional Type-I transitions. Zeeman slowing on a Type-II transition, though possible in principle, is difficult to realize~\cite{OspelkausDiatomicZeemanTheory, OspelkausTypeIIZeemanSlowing, lde2023}. Fortunately, these challenges have been successfully surmounted experimentally. In 1-D (e.g. for laser slowing), dark ground states are destabilized by rapidly switching the polarization of the slowing laser~\cite{DoyleSrOHMOT} or by applying a static magnetic field transverse to the slowing laser polarization~\cite{BarrySrFLaserSlowing}. In 3-D (e.g. in a MOT), destabilization is achieved by rapidly switching the laser polarizations~\cite{NorrgardSrF_rfMOT} or by applying two laser frequencies with orthogonal polarizations~\cite{CournolComparatDualPolarizationMOT, TarbuttSteimleDualFreqMOTTheory, HWilliamsCaFDualFreqMOT}. The destablization rate must be on the order of the scattering rate for effective application of optical forces. For numerical simulations of laser slowing in this work, we focus only on the application of a transverse static magnetic field for remixing, while for simulations of the MOT, we focus only on the use of two laser frequencies with orthogonal polarizations.

We propose using an ablation-loaded cryogenic buffer gas beam (CBGB) --- as is typically used in molecular laser cooling experiments~\cite{DeMilleSrFMOT, TarbuttCaFMOT, YeYOMOT, HutzlerCBGBReview} --- as a cold, bright source of atoms for a Group IV atomic laser cooling and trapping experiment. Typical experiments using CBGB sources report low initial velocities of $\sim\! 100 - 200\,\text{m/s}$~\cite{HutzlerCBGBReview}. Since laser slowing using a Type-II transition is inefficient compared to Type-I Zeeman slowers~\cite{BarrySrFLaserSlowing}, the low initial forward velocity of atoms in CBGBs is attractive. Moreover, ablation loading leads to high initial temperatures ($T \sim\! 4000\,\text{K}$) of the ablated species~\cite{StamperKurnRoomTempBGB}, which should naturally populate the metastable $^3\!P_1$ state that serves as the ground state of the optical cycling scheme. For example, we estimate that $\sim\! 30\%$ of the population of Sn atoms will be in the $^3\!P_1$ state. Existing data and theory calculations~\cite{CandSiCollisionsWithHe, SnCollisionalQuenching} suggest that the collisional quenching rate of metastable Group IV atoms in the buffer gas to the ground state is much slower than a typical extraction rate from the source cell.\footnote{If collisional quenching does turn out to be a problem, it would be possible to optically pump population into the $^3\!P_1$ state.}

In this work, we assume the source to have the same properties as the CBGB used in SrF laser cooling and trapping~\cite{BarrySrF_CBGBProperties}. The buffer gas is helium at $T \sim\! 4$ K, with flow rate $\mathcal{F} \sim\! 5$ sccm. The initial longitudinal velocity distribution is roughly Gaussian, with mean longitudinal velocity $\bar{v}_L = 140\,\text{m/s}$ and full width at half maximum (FWHM) $\Delta{v_L} = 75\,\text{m/s}$. The initial transverse velocity distribution is also roughly Gaussian, with mean $\bar{v}_T = 0\,\text{m/s}$ and FWHM $\Delta{v_T} = 75\,\text{m/s}$. We assume a beam brightness across all isotopes of $\mathcal{B} \sim\! 1.5\times 10^{12}$ atoms in the $^3\!P_1$ metastable state per steradian per pulse, commensurate with the measured brightness of Sr atoms in the SrF beam~\cite{VarunJorapurThesis} after accounting for the metastable state population. We remark that using a thermal beam source of atoms may also be viable, but we do not explore this path here.

\section{Numerical Simulations of Optical Forces}\label{sec:NumericalSimMethods}

In previous work from our group \cite{lde2023}, simulation software based on solving the optical Bloch equations (OBEs) was developed to investigate molecular laser cooling and trapping. In this work, we adapt this software for the case of Group IV atomic laser cooling and trapping. Details of the formalism can be found in Refs.~\cite{lde2023, DeMilleSrFCollisions}. In short, these simulations yield the acceleration of atoms due to optical forces.\footnote{The acceleration is averaged over optical scattering cycles and randomized laser phases.} Our simulations neglect particle-particle interactions and collisions.\footnote{This precludes direct estimation of MOT loss rate, atom number, density, or optical thickness.} MOTs realized using Type-II transitions are typically sparse enough that simulations under these assumptions still accurately capture the relevant physics of such MOTs.

To determine the force on atoms from external electromagnetic fields, we evolve the quantum state of the system until observables reach a periodic steady state \cite{TarbutOBESimulation}. We calculate the force using the Heisenberg equation of motion, averaged over one period of the Hamiltonian, and hence obtain the acceleration $a(r,v)$ due to the radiative force as a function of atom displacement from the MOT center, $r$, and velocity, $v$. 

From $a(r,v)$, we can evolve atomic trajectories in phase space. These trajectories incorporate stochastic photon recoils from spontaneous emission as atoms repeatedly absorb and emit photons from the cooling laser. Further details of our numerical simulation analysis are given in Appendix \ref{sec:AppxMOTSimDetails}.

\section{Capture into Red-Detuned MOT}\label{sec:CaptureMOT}

\begin{figure}
    \includegraphics[width=\linewidth]{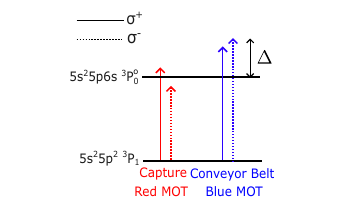}%
    \caption{Dual frequency/polarization mechanism used for capture and conveyor belt MOTs of Sn atoms. Both configurations consist of two co-propagating laser beams with orthogonal circular polarizations. $\Delta$ is the detuning of a laser frequency from the $^3\!P_1 \rightarrow {^3}\!P_0^\circ$ resonance.}
    \label{fig:MOTFig}
\end{figure}

We consider a dual-frequency MOT as the capture MOT stage for Sn. The dual-frequency mechanism employs one circularly polarized laser beam red-detuned to the optical cycling transition, and a second, blue-detuned, co-propagating laser beam with orthogonal polarization (See Fig.~\ref{fig:MOTFig}). This generates an appreciable spatially restoring force on a Type-II transition~\cite{TarbuttSteimleDualFreqMOTTheory, TarbuttDevlinTypeIITheory}. 

The optimized and experimentally viable parameters for this stage are shown in Table \ref{tab:RedMOT_parameters}. With Gaussian laser beams of $w_\text{MOT} = 7$ mm ($1/e^2$ intensity radius), the scheme requires a power $P_\text{beam} \sim\! 200\,\text{mW}$ per beam of the MOT. This can be achieved with a total of $P_\text{tot}\approx 200\,\text{mW}$ by using a single-beam configuration as in Ref.~\cite{DeMilleSrFODT, DeMilleSrFCollisions}. The axial magnetic (B-)field gradient is $\partial B/\partial z = 50\,\text{G/cm}$.

\begin{table*}
\caption{Parameters used for Sn capture and compressed MOT simulations (see 
Fig.~\ref{fig: red MOT heat map} and Fig.~\ref{fig: compression sequence}). For the compression stage, the parameters shown are those for the last stage of compression. Laser power $P_i$, saturation parameter $s_i$, and detuning $\Delta_i$ are given for each polarization 
component $\hat{p}_i$ (for the given polarization, we assume the beam wave vector $\hat{k}_i$ points towards $-\hat{x}$ and that the B-field increases along $+\hat{x}$) in the dual frequency/polarization scheme. $\partial B/\partial z$ is the axial B-field gradient. The Gaussian laser beam size ($1/e^2$ intensity radius) is $w_\text{MOT} = 7$ mm in both MOT stages. In the results, $v_{\rm cap}$ is the capture velocity; $T$ and $\sigma$ are the temperature and cloud size; $P_{\rm exc}$ is the average population in the excited state.}
\label{tab:RedMOT_parameters}
\begin{ruledtabular}
\begin{tabular}{c c c c c c c c c c}
\multirow{2}{*}{MOT Stage}
  & \multirow{2}{*}{$\frac{\partial B}{\partial z}$ (G/cm)}
  & \multicolumn{4}{c}{Laser Parameters}
  & \multicolumn{4}{c}{Results} \\
\cline{3-6} \cline{7-10}
  & 
  & $P_i$ (mW) & $s_i$  & $\Delta_i$ ($\Gamma$) & $\hat p_i$
  & $v_{\rm cap}$ (m/s) & $T$ (mK) & $\sigma$ (mm) & $P_{\rm exc}$ (\%) \\
\hline
\multirow{2}{*}{Capture}
  & \multirow{2}{*}{50}
  & 172 & 1.5 & $-1.0$ & $\sigma^{-}$
  & \multirow{2}{*}{28.5} & \multirow{2}{*}{225} & \multirow{2}{*}{1.9} & \multirow{2}{*}{7.8} \\
  & 
  &  34 & 0.3 & $+0.5$ & $\sigma^{+}$ &  &  &  & \\
\hline
\multirow{2}{*}{Compress}
  & \multirow{2}{*}{150}
  &   5 & 0.05 & $-0.5$ & $\sigma^{-}$
  & \multirow{2}{*}{---}  & \multirow{2}{*}{8}   & \multirow{2}{*}{0.5} & \multirow{2}{*}{0.6} \\
  & 
  &   1 & 0.01 & $+0.25$ & $\sigma^{+}$ &  &  &  & \\
\end{tabular}
\end{ruledtabular}
\end{table*}

\begin{figure}
    \includegraphics[width=\linewidth]{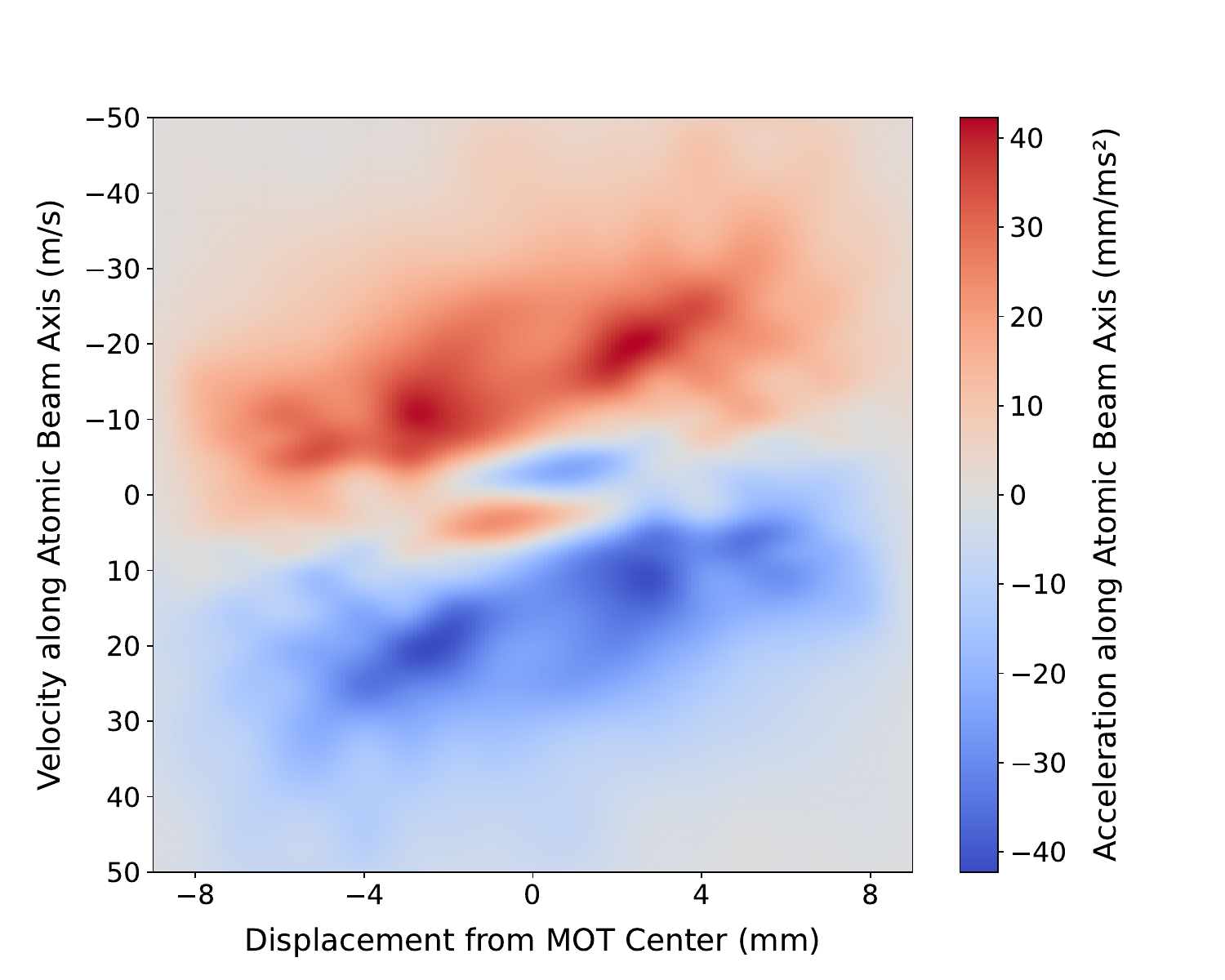}%
    \caption{Acceleration experienced by Sn atoms in the capture MOT as a function of velocity and displacement towards the MOT center, orthogonal to the MOT's magnetic quadrupole field symmetry axis, for the parameters listed in Table \ref{tab:RedMOT_parameters}. Note the strong sub-Doppler heating at low velocities near the trap center, as seen from the reversed sign of acceleration in this region. This leads to a high equilibrium temperature in the MOT.}
    \label{fig: red MOT heat map}
\end{figure}

A map of the acceleration experienced by Sn atoms in the capture MOT is shown in Fig.~\ref{fig: red MOT heat map}. The plot clearly indicates the presence of strong Doppler cooling at large velocities, for example with peak deceleration $a \approx -40$ mm/ms$^2$ at $v\approx  20$ m/s and $r\approx -4$ mm. At lower velocities, $v \lesssim 5\,\text{m/s}$, the acceleration changes sign, indicating the presence of sub-Doppler heating. This behavior is typical for Type-II transitions~\cite{TarbuttDevlinTypeIITheory}, and can be more clearly seen in plots of the spatial and velocity-averaged forces derived from integrating $a(r,v)$ over $v$ and $r$, respectively, as shown in Appendix \ref{sec:AppxMOTSimDetails}.

To determine capture velocity, we initialize atoms at $r = -14$ mm, with increasing velocities $v$.  An atom is deemed captured if, after $t = 50$ ms of evolution, its trajectory has converged onto a steady-state oscillation in phase space, with the spatial extent of that oscillation $r_{\rm max} < w_{\rm MOT}$. We find that this condition is met for all trajectories for which the velocity reverses sign before the displacement along $+\hat{r}$ exceeds $w_{\rm{MOT}} = 7$ mm. We find a capture velocity of $v_c \approx 28.5$ m/s. This value is 2-3 times higher than what is experimentally measured for laser-cooled molecules like CaF and SrF~\cite{HWilliamsCaFDualFreqMOT, SteineckerSrFThesis}, due to the larger natural linewidth of the cooling transition and $\sim\! 2\times$ higher photon momentum in Sn. Plots of trajectories for determining $v_c$ are shown in Appendix \ref{sec:AppxMOTSimDetails}. 

To characterize the temperature and size of the MOT, we extend the trajectory of captured and trapped atoms by another $t = 50$ ms to allow MOT observables to reach equilibrium. Using these trajectories, we calculate the rms displacement $\sigma$ (which we denote the cloud size), velocity $v_\sigma$, and temperature $T = mv_\sigma^2 / k_B$, by averaging the position and velocity over the last 10 ms. The final size and temperature are averaged over ten such trajectories. For the capture MOT stage, we find $T \sim\! 225\,\text{mK}$ and $\sigma \sim\! 1.9\,\text{mm}$. For context, the Doppler temperature for this cooling transition in Sn is $T_D = 760\,\mu\text{K}$. We attribute the extremely hot temperature and large size of the capture red MOT to the large natural linewidth and Type-II nature of the optical cycling transition; a similarly hot MOT was observed in Ref.~\cite{TruppeAlFMOT}. A typical in-MOT trajectory is shown in Appendix \ref{sec:AppxMOTSimDetails}.

We note the possibility of two-photon ionization (TPI) when using the $^3\!P_1 \rightarrow {^3}\!P_0^\circ$ optical cycling transition in Sn. This would manifest as one-body loss of the form $dN/dt = -\alpha N$. The loss rate due to TPI, $\alpha$, is given by:
\begin{equation}
\alpha = \frac{\sigma P_\text{exc} I_\text{tot}}{\hbar\omega},
\label{eq:TPI}
\end{equation}
where $\sigma$ is the TPI cross section, $P_\text{exc}$ is average population in the excited state, $I_\text{tot}$ is total laser intensity from all six passes, and $\omega$ is the angular frequency of the cycling transition. According to Ref.~\cite{saloman1992resonance}, Sn was calculated to have $\sigma \sim\! 6\times 10^{-19}\,\text{cm}^2$ for this transition. Using the parameters in Table \ref{tab:RedMOT_parameters}, we find $\alpha \sim\! 0.1\,\text{s}^{-1}$, leading to a lifetime $\tau \sim\! 10$ s in the capture MOT. For applications where the cold atoms are subsequently loaded into a far-detuned optical trap, this lifetime is not a limiting factor.

\section{White Light Slowing}
For producing atoms with sufficiently low velocity to be captured in the MOT, we consider here the method of white light slowing (WLS) applied to the CBGB beam of atomic Sn. In WLS, the slowing laser spectrum is frequency-broadened to simultaneously cover many velocity classes. WLS is used in many molecular laser cooling experiments, enabling trapping of up to $\sim\! 10^4$ molecules in a MOT~\cite{DeMilleSrFCollisions, DoyleSrOHMOT}. 

In this section, we use simulations to study WLS of a CBGB of Sn from the source region to the MOT loading region. The slowing laser detuning $\Delta_s$ is set roughly to resonance with the Doppler-shifted line of atoms at the center of the velocity range we wish to address. The laser spectrum is broadened using electro-optic modulators (EOMs). The sideband frequency $\Omega \sim\! 1\Gamma$ is chosen to address as broad a range of velocities as possible. Commercial resonant-frequency EOMs for UV laser wavelengths can provide only modest modulation depth $\beta$, so we set $\beta \leq 1.5\,\text{rad}$ in our WLS simulation. To achieve sufficient spectral width with such limited values of $\beta$, we send the slowing laser through two successive EOMs with different frequencies and modulation depths. Finally, we use a converging slowing laser beam, which provides strongest deceleration near the CBGB source and a small transverse confining force, both desirable for efficient delivery of slowed atoms to the MOT region~\cite{SteineckerSrFMOTCharacterization}.

We define a geometry where the longitudinal direction of the atomic beam is the $z$ axis. We compute the longitudinal acceleration $a_z(v_z, r, z)$ of an atom as a function of forward longitudinal velocity $v_z$, transverse displacement from the central beamline axis $r$, and longitudinal distance from the beam source $z$. For our converging slowing beam, we take the transverse deceleration to be $a_\perp = a_z\tan\theta'$. The slowing laser polarization is linear and defines the $x$-axis. To remix the dark ground states, we apply a transverse static magnetic field $\vec{B} = B_0(\cos \theta\, \hat{x} + \sin \theta \, \hat{y})$, with $B_0 = 15\,\text{G}$ and $\theta = \arccos(1/\sqrt{3}) = 54.7\degree$.\footnote{Only the relative angle between the $B$-field and laser polarization matters. In a real experiment, it may be easier to apply $\vec{B}$ and rotate the laser polarization by $\theta = 54.7\degree$ using a half-wave plate.} The value of $B_0$ corresponds to a Larmor precession rate $g\mu_BB_0 \approx \hbar\Gamma$, where $g = 1.502$ is the $g$-factor of the $^3\!P_1$ state~\cite{NIST_ASD}. The value of $\theta$ provides the maximal remixing rate in a $J=1 \rightarrow J'=0$ system~\cite{BerkelandBoshierDarkStateRemix}. 

The WLS parameters that we determined to provide effective slowing of Sn atoms for capture into the MOT are shown in Table \ref{tab:WLS parameters}. The spectral profile of slowing light is shown in Fig.~\ref{fig:Sn WLS spectral profile}, and a plot of the atomic deceleration is shown in Appendix \ref{sec:AppxWLSSimDetails}. The short excited-state lifetime of the optical cycling transition in atomic Sn makes it possible to apply a high scattering force to the atoms. We find that a slowing distance of $z_\text{MOT} = 500\,\text{mm}$ is sufficient for effective WLS. This is shorter than in typical molecular MOT setups~\cite{TarbuttCaFMOT, YeYOMOT, YanBaFMOT}, resulting in a larger solid angle subtended by the MOT region and hence greater slowing efficiency~\cite{lde2023}. 

\begin{table}
\caption{Slowing laser parameters used for simulations of white light slowing of Sn atoms. We assume a converging slowing beam with $w = 7.5$ mm ($1/e^2$ intensity radius) at the MOT center ($z = 500$ mm), half-angle $\theta' = 4.5$ mrad, and linear polarization. $P$ indicates the slowing laser power, $\bar{s}$ the saturation parameter spatially averaged along the slowing beamline axis, $\Delta_s$ the slowing laser carrier frequency detuning, $\Omega_i$ the modulation frequency of each EOM, and $\beta_i$ the modulation depth of each EOM.}
\label{tab:WLS parameters}
\begin{ruledtabular}
\begin{tabular}{c c c c c c c c}
$P$ (mW) & $\bar{s}$ & $\Delta_s$ ($\Gamma$) & $\Omega_1$ ($\Gamma$) & $\Omega_2$ ($\Gamma$) & $\beta_1$ (rad) & $\beta_2$ (rad) \\
\hline
300 & 3.3 & $-11.75$ & 1.5 & 3.4 & 1.5 & 1.5\\
\end{tabular}
\end{ruledtabular}
\end{table}

\begin{figure}
    \includegraphics[width=\linewidth]{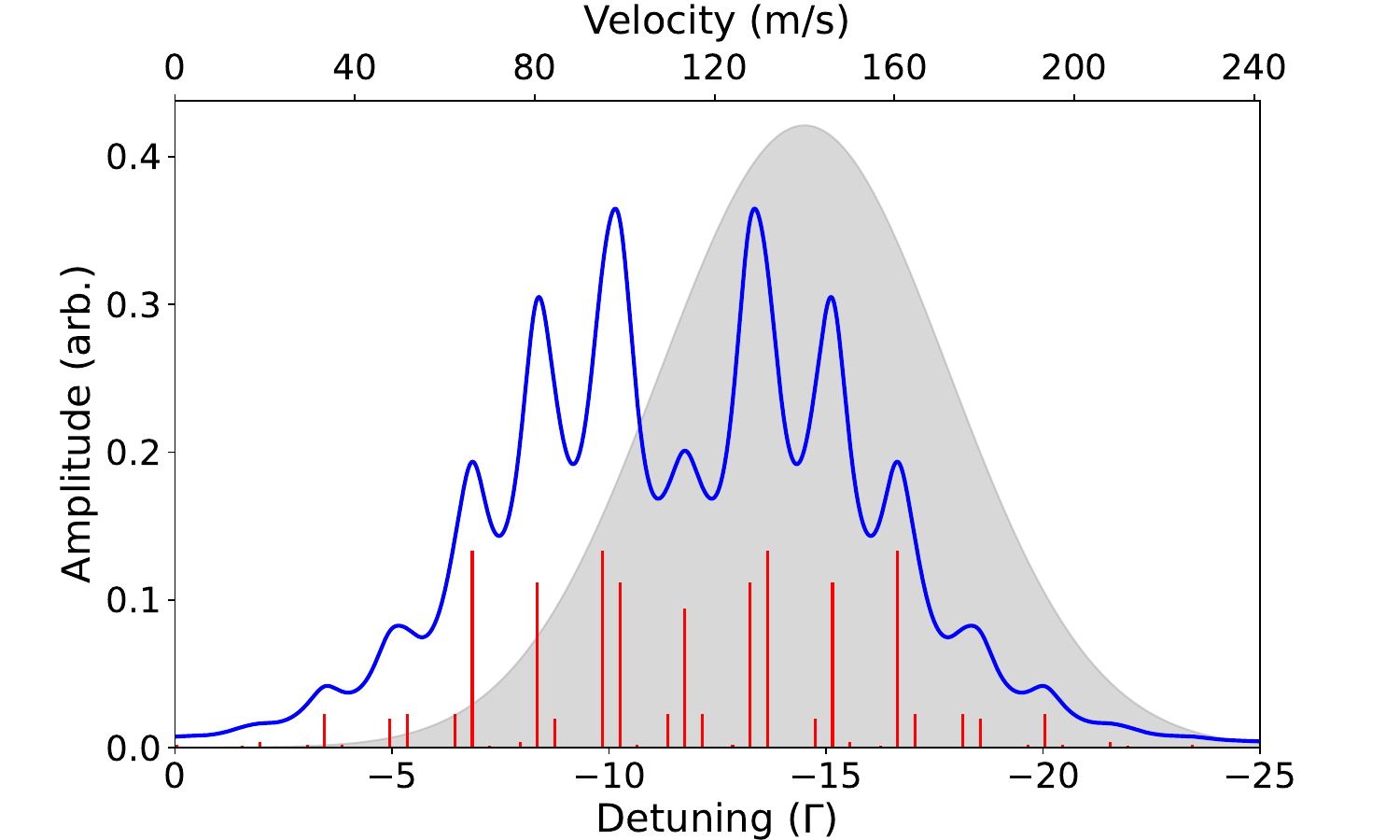}
    \caption{Spectral profile of slowing light. The red lines show the laser spectrum. The blue curve shows a convolution of this spectrum with a Lorentzian of linewidth $\Gamma$. The shaded gray region shows the initial longitudinal velocity distribution of Sn atoms from the CBGB, with the scale chosen such that the velocity and Doppler-shifted atomic resonance frequency coincide. The laser parameters are given in Table \ref{tab:WLS parameters}.}
    \label{fig:Sn WLS spectral profile}
\end{figure}

To find the capture efficiency of our WLS scheme, we perform simulations of atom trajectories undergoing WLS. Collisions of the atoms with He buffer gas immediately after exiting the source cell significantly affect CBGB properties ~\cite{BarrySrF_CBGBProperties}. To account for this, we follow the procedure used in Ref.~\cite{lde2023} to obtain an initial position distribution for the atoms: we convolve an initial uniform disk distribution (radius $R = 1.5$ mm, the source cell aperture size) with a Gaussian distribution of FWHM $\Delta v_T \times z_Q/v_L$, where $z_Q = 7.5$ mm is the zone of freezing (after which collisions between atoms and He are unlikely). We draw the initial transverse position of atoms at $z = z_Q$ from this convolved distribution. We then take the initial velocity distributions of Sn atoms at $z = z_Q$ to be the same as those of the SrF molecular laser cooling experiment (see Sec.~\ref{sec:LaserCoolingSchemeSpinZero}). Using these initial conditions, we run $1\times 10^8$ trajectory simulations. We solve the differential equations $\ddot{z} = a_z(\dot{z}, r, z)$, $\ddot{x} = a_\perp x/r$, and $\ddot{y} = a_\perp y/r$, where the dot indicates a time derivative and $a_z$ is obtained using the parameters in Table~\ref{tab:WLS parameters}, to propagate the atom trajectory from $z = z_Q$ to MOT capture region. We assume that an atom is captured by the MOT if it simultaneously satisfies two conditions \cite{lde2023}:
\begin{enumerate}
    \item $v_z < v_\text{cap}$ at the location of the MOT ($z = z_\text{MOT}$).
    \item $r < w_\text{MOT}$ at the location of the MOT. In other words, we take the MOT capture radius to be defined by the MOT laser beam radius.
\end{enumerate}

\begin{figure}
    \includegraphics[width=\linewidth]{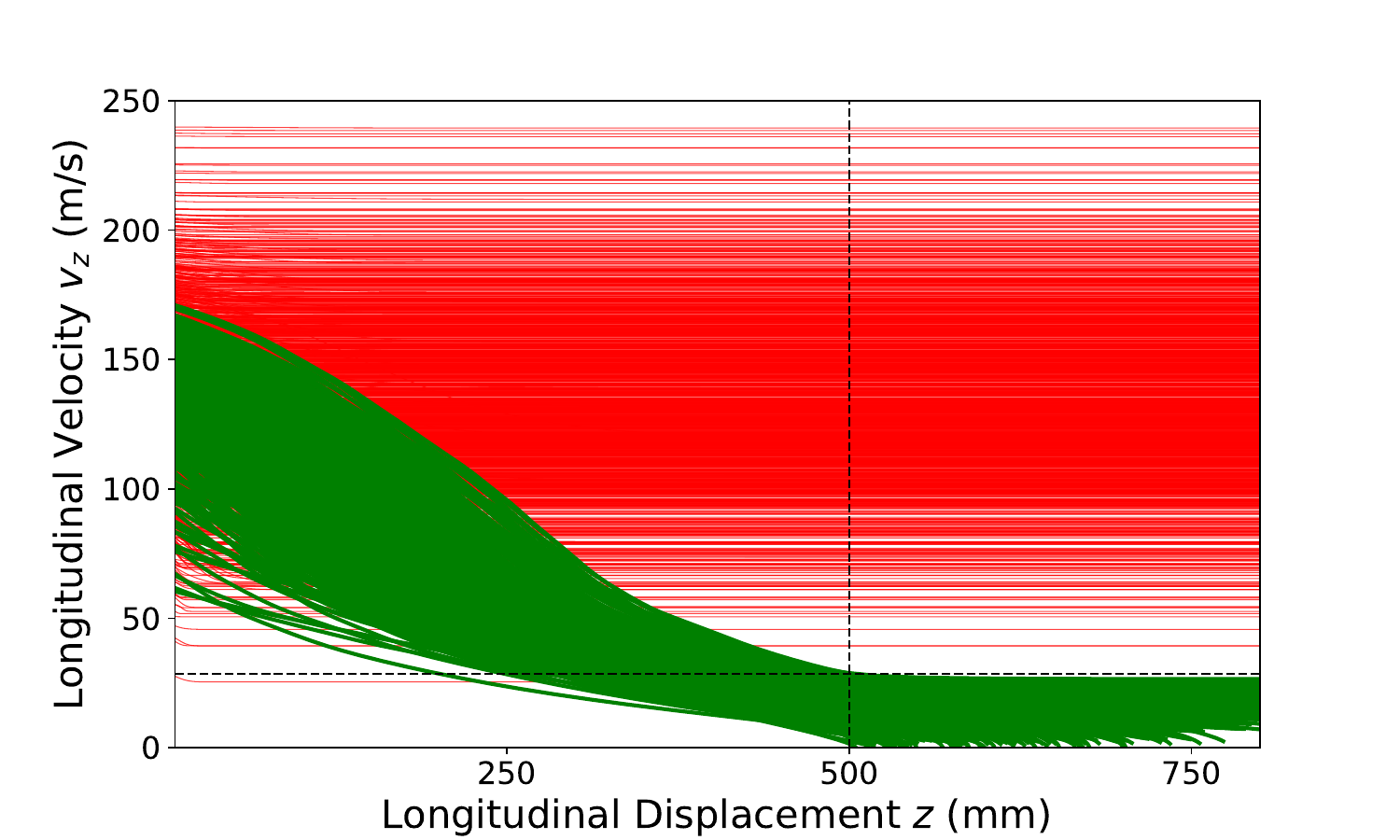}%
    \caption{Monte Carlo simulation of Sn atom trajectories propagating from the CBGB source to the MOT region under the influence of WLS, with parameters as described in the text. Only the longitudinal direction is shown. Trajectories in green (only every 20$^{\rm th}$ trajectory is shown)  indicate successful capture. Trajectories in red (only every 80000$^{\rm th}$ trajectory is shown) indicate failure to capture. The computed efficiency of capture is $\sim\!1\times 10^{-4}$.}
    \label{fig:AtomTrajLongitudinal}
\end{figure}

The result of this trajectory simulation is shown in Fig.~\ref{fig:AtomTrajLongitudinal}. We find that the fraction of atoms captured is $\sim\! 1 \times 10^{-4}$. With our assumed Sn CBGB brightness and for the case of $^{120}$Sn (the most abundant isotope), this implies capture of $\sim\! 10^7$ Sn atoms in the MOT. Note that similar efficiency calculations were performed in Ref.~\cite{lde2023} for WLS of SrF molecules, where a capture efficiency over $100\times$ smaller was estimated.

\section{Further Cooling and Trapping}

\subsection{Compressed MOT}
The parameters used for the Sn capture red MOT were chosen to maximize the capture velocity. This requires high scattering rate and leads to high temperature and large size, due to use of a Type-II cycling transition. To further cool the captured atoms, we propose to apply a compressed MOT (cMOT) stage after the capture MOT. This enables efficient transfer of cold atoms into subsequent cooling stages for direct loading of a conservative trap.

\begin{figure}
    \centering
    \includegraphics[width=\linewidth]{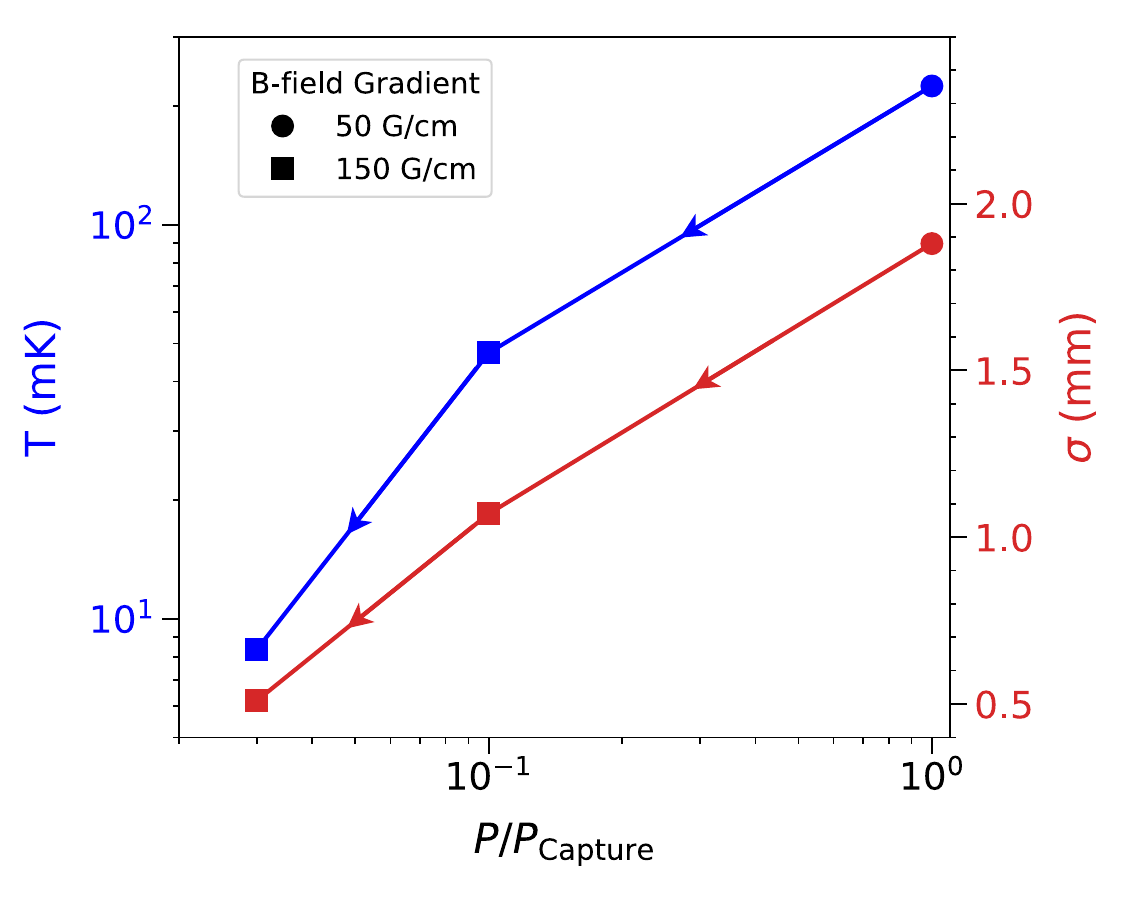}
    \caption{Simulated compression sequence for the Sn red MOT. Arrows indicate the progression of the sequence. The temperature $T$ and cloud size $\sigma$ are reduced by factors of 30 and 4, respectively, as laser power $P$ is reduced ($P_\text{capture}$ is the laser power used for capture MOT stage), laser detuning is decreased, and B-field gradient is increased. We anticipate a compression sequence time of $\sim\! 25$ ms, commensurate with what is typically used in MOT experiments~\cite{DeMilleSrFCollisions}.}
    \label{fig: compression sequence}
\end{figure}

The cMOT is implemented by reducing the intensity and increasing the B-field gradient. The reduced scattering rate and higher B-field gradient drastically lowers the equilibrium temperature and tightens the spatial extent of the MOT~\cite{CornellCMOTBehavior}. We use our simulations to find experimentally viable parameters for the cMOT sequence (see Table~\ref{tab:RedMOT_parameters}). The compression sequence is shown in Fig.~\ref{fig: compression sequence}. 

We find that decreasing the laser intensity to $3\%$ of its value in the capture MOT, tripling the axial B-field gradient $\partial B/\partial z$, and reducing the laser detunings $\Delta_i$ of both polarization components leads to a Sn cMOT with temperature $T \sim\! 10\,\text{mK}$ and size $\sigma \sim\! 500\,\mu\text{m}$. The average excited state population decreases from $\sim8\%$ to $\sim 1\%$ during the compression sequence, leading to reduced heating from photon scattering. In total, the phase space density (PSD) increases by nearly five orders in magnitude from the capture MOT to the cMOT.

\subsection{Conveyor Belt Blue-Detuned MOT}

We next consider further cooling of the Sn cloud using the ``conveyor belt" blue-detuned MOT (CB MOT) technique \cite{lhd2025}. A conventional blue MOT provides sub-Doppler cooling while maintaining spatial confinement in systems with Type-II transitions \cite{jdt2018}. First demonstrated on the D$_2$ line in $^{87}$Rb atoms \cite{jdt2018}, the blue MOT has also been successfully applied to a variety of laser-coolable molecules \cite{bay2023,DeMilleSrFCollisions,hlc2024,CaOH_CBMOT}.

The CB MOT is a novel variant of the blue MOT scheme that was recently theoretically proposed \cite{lhd2025} and experimentally demonstrated in laser-cooled CaOH, CaF, BaF, and SrOH molecules \cite{CaOH_CBMOT, CaF_CBMOT, zeng2025bafbluemot_published,SrOHODTandCBMOT}. This new scheme provides an additional position-dependent trapping force, which enables formation of a more compressed atomic cloud compared to a conventional blue MOT. The CB MOT uses two blue-detuned laser beams with orthogonal polarizations and closely spaced ($< 1\Gamma$) frequencies (see Fig.~\ref{fig:MOTFig}).

\begin{figure}
    \centering
    \includegraphics[width=\linewidth]{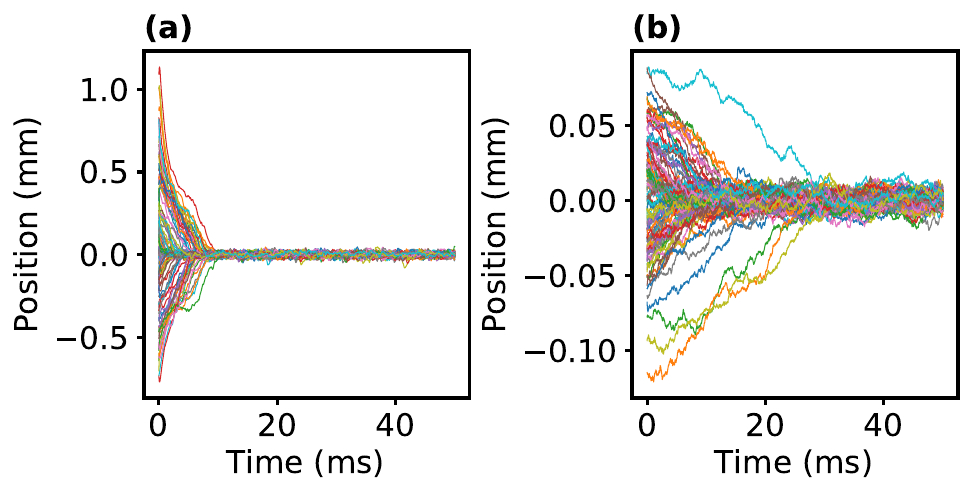}
    \caption{Simulations of the Sn CB MOT. Each figure shows 100 atomic trajectories with initial positions and velocities randomly sampled from the distribution after the cMOT. All relevant parameters are in Table \ref{tab:CBMOTMOT_parameters}. (a) First stage CB MOT. With an initial temperature $ T \sim 10\,\text{mK}$ and cloud size $\sigma \sim 500\,\mu\text{m}$ from the cMOT, this stage bring atoms to $T \sim\! 75\ \mu$K and $\sigma \sim\! 20\,\mu\text{m}$ after $t \sim\! 10$ ms. (b) Second stage CB MOT. This bring atoms to $T \sim\! 15\, \mu\text{K}$ and $\sigma \sim\! 10\,\mu\text{m}$ after $t \sim\! 15$ ms.}
    \label{fig: CBMOT_Compression}
\end{figure}

\begin{table*}
\caption{Parameters used for Sn CB MOT simulations (See Fig.~\ref{fig: CBMOT_Compression}). Laser power $P_i$, saturation parameter $s_i$, and detuning $\Delta_i$ are given for each polarization component $\hat{p}_i$ (for the given polarization, we assume same beam wave vector $\hat{k}_i$ direction as in previous MOT stages, see Table \ref{tab:RedMOT_parameters}) in the dual frequency scheme. $\partial B/\partial z$ is the axial B-field gradient. The Gaussian laser beam size ($1/e^2$ intensity radius) is $w_\text{MOT} = 7$ mm in both stages. $t_D$ is the compression time; $T$ and $\sigma$ are the temperature and cloud size.}
\label{tab:CBMOTMOT_parameters}
\begin{ruledtabular}
\begin{tabular}{c c c c c c c c c}
\multirow{2}{*}{Label} & \multirow{2}{*}{$\frac{\partial B}{\partial z}$ (G/cm)} & \multicolumn{4}{c}{Laser Parameters} & \multicolumn{3}{c}{Results} \\
\cline{3-6} \cline{7-9}
& & $P_i$ (mW) & $s_i$ & $\Delta_i(\Gamma)$ & $\hat{p}_i$ & $t_D$ (ms) & $T$ ($\mu\text{K}$) & $\sigma$ ($\mu\text{m}$) \\
\hline
\multirow{2}{*}{Stage 1} 
& \multirow{2}{*}{30} & 172 & 1.5 & $+3.2$ & $\sigma^-$ & \multirow{2}{*}{10} & \multirow{2}{*}{75} & \multirow{2}{*}{20} \\
& & 172 & 1.5 & $+3.0$ & $\sigma^+$ & & & \\
\hline
\multirow{2}{*}{Stage 2} 
& \multirow{2}{*}{40} & 114 & 1.0 & $+4.2$ & $\sigma^-$ & \multirow{2}{*}{15} & \multirow{2}{*}{15} & \multirow{2}{*}{10} \\
& & 114 & 1.0 & $+4.0$ & $\sigma^+$ & & & \\
\end{tabular}
\end{ruledtabular}
\end{table*}

For the CB MOT of Sn atoms, we found a two-stage scheme to be effective. First, we use a relatively high laser intensity and low B-field gradient to maximize the number of atoms transferred from the cMOT to the CB MOT. Next, we reduce the laser intensity and increase the B-field gradient and laser detuning to further cool and compress the atomic cloud. We initialize atomic positions and velocities by randomly sampling from a distribution corresponding to the final size and temperature after the cMOT stage. Using the same method introduced in Section~\ref{sec:CaptureMOT}, we then simulate the atomic compression trajectories over a 50 ms duration. The relevant experimental parameters used in each CB MOT stage are shown in Table \ref{tab:CBMOTMOT_parameters}. 

Fig.~\ref{fig: CBMOT_Compression} shows atomic trajectories during each stage of the process. At the end of the first stage, which lasts $t \sim\! 10\,\text{ms}$, we find an equilibrium temperature $T \sim\! 75\,\mu\text{K}$ and cloud size $\sigma \sim\! 20\,\mu\text{m}$. During the second stage, which lasts an additional $t \sim\! 15\,\text{ms}$, the cloud is further cooled and compressed to $T \sim\! 15\,\mu\text{K}$ and $\sigma \sim\! 10\,\mu\text{m}$. In a real experiment, the atomic cloud will likely be hotter and larger than what was found in this simulation, due to our use of a semi-classical model and effects we neglected, like photon re-scattering and experimental imperfections. However, we do not expect the deviation of simulation from reality to be substantial \cite{CaOH_CBMOT}.

\section{Outlook}

\subsection{Proposed Experimental Setup}

Here, we briefly discuss a proposed experimental setup for realizing a Sn MOT, synthesizing our numerical simulation results with realistic experimental constraints. Commercially available laser systems at $\lambda = 303\,\text{nm}$ can provide up to $P \approx 1\,\text{W}$ of power. We anticipate using acousto-optic modulators (AOM) to generate the laser frequencies needed for WLS and all MOT stages. Typical UV AOMs in a double-pass configuration have peak diffraction efficiency of $\sim \!45\%$ due to the strong polarization preference of UV AOM crystals. 

In laser cooling of SrF molecules, WLS and capture MOT light are typically simultaneously on for optimal capture into the MOT. Since our results indicate $P \sim\! 300\,\text{mW}$ is needed for efficient WLS of Sn atoms, we restrict the capture MOT power budget to $P \sim\! 300\,\text{mW}$. Once capture is complete, WLS light can be shut off and used for subsequent MOT stages. Therefore, we restrict the CB MOT power budget to $P \sim\! 450\,\text{mW}$. All told, this means the experiment will require a single-beam-pass MOT, where laser power is recycled through all six passes of the chamber. While not ideal for beam pointing stability and alignment, this type of beam delivery was routinely and successfully used in older-generation molecular laser cooling experiments for similar reasons~\cite{DeMilleSrFODT, DeMilleSrFCollisions}. 

\subsection{Prospects for Further Cooling}
Various free-space cooling methods have achieved even lower temperatures in atoms and molecules. One approach --- velocity-selective coherent population trapping (VSCPT)~\cite{aac1988} and $\Lambda$-enhanced gray molasses~\cite{CheukAndereggLambdaImaging,gfs2013,DeMilleSrFODT} --- uses coherent, low-momentum dark states in a $\Lambda$-system that decouple from the laser field. The formation of such states enables atoms or molecules to reach temperatures well below the temperature reached in our CB MOT simulation. Experiments with $^4$He used Zeeman sublevels ($\ket{J=1, m=\pm 1}$ and $ \ket{J’=1, m=0}$) to form a $\Lambda$ structure and obtain sub-recoil temperatures~\cite{3DVSCPT}. For a $J=1 \rightarrow J’=0$ configuration, theoretical work predicts that VSCPT should also be feasible~\cite{3d3+1VSCPT}. Therefore, we expect Sn atoms in a VSCPT scheme to theoretically achieve similar performance as $^4$He atoms, with temperatures approaching the recoil limit ($\sim\! 1\,\mu\text{K}$). Additionally, since potential precision measurement experiments with Sn involve loading atoms into an optical lattice, we anticipate that degenerate Raman sideband cooling could be used for further cooling ~\cite{DRamanSidebandCooling_Steven,huang2021dark}.

\section{Summary}
We have presented a realistic experimental scheme for laser cooling and trapping of Sn atoms.  Based on the closely analogous structure of other Group IV elements (in particular, Si, Ge, and Pb), similar schemes should be effective in cooling and trapping these species as well. In forthcoming papers, we will explore the use of laser-cooled Group IV atoms for precision measurements and tests of fundamental physics, including atomic parity violation.

\begin{acknowledgements}
We thank Silviu M. Udrescu for discussions about the applications of laser-cooled tin atoms for precision measurement. G.Z. and Q.W. acknowledge support from the Air Force Office of Scientific Research (AFOSR MURI Grant No. FA9550-21-1-0069). M.V. acknowledges support from the Natural Sciences and Engineering Research Council of Canada (NSERC). 
\end{acknowledgements}

\section*{Data Availability}
The data that support the findings of this article are openly available \cite{dataset_forpaper}.

\appendix

\section{Calculating the Metastable $s^2p^2\,{^3}\!P_1$ State Lifetime}\label{sec:AppxMetastableLifetime}

The ground state of the proposed optical cycling transition for Group IV elements is the metastable $s^2 p^2 \, {^3}\!P_1$ state. This state only has one decay path, which goes to the absolute ground state ($s^2p^2\,{^3}\!P_1 \leftrightarrow s^2p^2\,{^3}\!P_0$) via a pure M1 transition. Here, we analytically estimate the natural lifetime of the $s^2p^2\,{^3}\!P_1$ state to verify that the lifetime is long compared to typical magneto-optical trap (MOT) timescales. Throughout, we work in the $LS$-basis, which is approximately good for the lighter Group IV elements (C, Si, Ge) but requires some correction for the heavier Group IV elements (Sn, Pb). We compare our estimate to theoretical calculations done for C, Si, Sn, and Pb.

The magnetic dipole (M1) operator is given by: \begin{equation}
    \hat{\mu} = -\frac{\mu_B}{\hbar}(g_L \mathbf{L} + g_S\mathbf{S}).
\end{equation} 
We take $g_L \approx 1$ and $g_S \approx 2$. The spontaneous decay rate from $\ket{e} = \ket{J',L',S'}$ to $\ket{g} = \ket{J,L,S}$ (transition energy $\omega_0$) is given by:
\begin{equation}
    \Gamma_{\rm M1} = \frac{\omega_0^3}{3\pi\varepsilon_0\hbar c^5\!}\frac{|\rmel{g}{\hat{\mu}}{e}|^2}{2J'+1}.
\end{equation}
We use standard results from angular momentum algebra and find:
\begin{widetext}
\begin{equation}
    \rmel{L,S,J}{\mathbf{L}}{L',S',J'} = \delta_{S,S'}(-1)^{L'+S+J'+1}\sqrt{(2J+1)(2J'+1)}\begin{Bmatrix}L' & J' & S \\ J & L & 1 \end{Bmatrix} \rmel{L}{\mathbf{L}}{L'},
\end{equation}
and
\begin{equation}
    \rmel{L,S,J}{\mathbf{S}}{L',S',J'} = \delta_{L,L'}(-1)^{L+S'+J+1}\sqrt{(2J+1)(2J'+1)}\begin{Bmatrix}S' & J' & L \\ J & S & 1 \end{Bmatrix} \rmel{S}{\mathbf{S}}{S'}.
\end{equation}

Using $\rmel{L}{\mathbf{L}}{L'} = \delta_{L,L'}\sqrt{L(L+1)(2L+1)}$, and likewise for $\mathbf{S}$, we substitute and obtain:

\begin{equation}
\Gamma_{\mathrm{M1}}
= \frac{\mu_B^2 \omega_0^3}{3\pi \varepsilon_0 \hbar c^5\!}\,
\frac{|\mathcal{M}|^2}{2J'+1},
\end{equation}
where
\begin{equation}
\begin{aligned}
\mathcal{M} &= 
\delta_{S,S'} \delta_{L,L'} (-1)^{L'+S+J'+1} \sqrt{(2J+1)(2J'+1)}
\sqrt{L(L+1)(2L+1)}
\begin{Bmatrix}
L' & J' & S \\
J  & L  & 1
\end{Bmatrix} \\
&\quad
+ 2\,\delta_{L,L'} \delta_{S,S'} (-1)^{L+S'+J+1} \sqrt{(2J+1)(2J'+1)}
\sqrt{S(S+1)(2S+1)}
\begin{Bmatrix}
S' & J' & L \\
J  & S  & 1
\end{Bmatrix}.
\end{aligned}
\end{equation}

\end{widetext}

For our transition of interest, we have $\ket{g} = \ket{0,1,1}$ and $\ket{e} = \ket{1,1,1}$. Therefore, $\rmel{^3\!P_0}{\hat{\mu}}{^3\!P_1} = \sqrt{2}\,\mu_B$ in the $LS$-basis. For the heavier elements, we find from Ref.~\cite{BiemontAstrophysForbiddenTransition} that $\rmel{^3\!P_0}{\hat{\mu}}{^3\!P_1} = 1.430\,\mu_B$ in Sn and from Ref.~\cite{SafronovaPbRMECalc} that $\rmel{^3\!P_0}{\hat{\mu}}{^3\!P_1} = 1.293\,\mu_B$ in Pb. We compile our results for all Group IV elements in Table \ref{tab:M1_transitions}, where we use the $LS$-coupling calculated values for C, Si, Ge and the literature values for Sn, Pb. Our estimation for C and Si matches well with the theoretically calculated result in literature~\cite{NIST_ASD}.

\begin{table*}[t]
\caption{Parameters of interest for the M1 transition $s^2p^2\, {^3}\!P_1 \leftrightarrow s^2p^2\, {^3}\!P_0$ for all Group IV elements. $k$ denotes transition wavenumber, $\omega_0$ denotes transition energy, $\Gamma_{\rm M1}$ denotes natural linewidth, and $\tau_g$ denotes lifetime of the metastable state.}
\label{tab:M1_transitions}
\begin{ruledtabular}
\begin{tabular}{c c c c c}
Element &
$k$ (cm$^{-1}$) &
$\omega_0 / 2\pi$ (THz) &
$\Gamma_{\mathrm{M1}} / 2\pi$ (Hz) &
$\tau_g$ (s) \\
\hline
Carbon (C)      & 16.42    & 0.49   & $1.25 \times 10^{-8}$ & $1.27 \times 10^7$ \\
Silicon (Si)    & 77.12    & 2.31   & $1.31 \times 10^{-6}$ & $1.22 \times 10^5\!$ \\
Germanium (Ge)  & 557.13   & 16.70  & $4.95 \times 10^{-4}$ & 322 \\
Tin (Sn)        & 1691.81  & 50.72  & 0.014                 & 11.48 \\
Lead (Pb)       & 7819.26  & 234.42 & 1.145                  & 0.14 \\
\end{tabular}
\end{ruledtabular}
\end{table*}

\section{Further Details on Atomic Tin MOT Simulations}\label{sec:AppxMOTSimDetails}
\begin{figure*}[t]
    \includegraphics[width=0.9\linewidth]{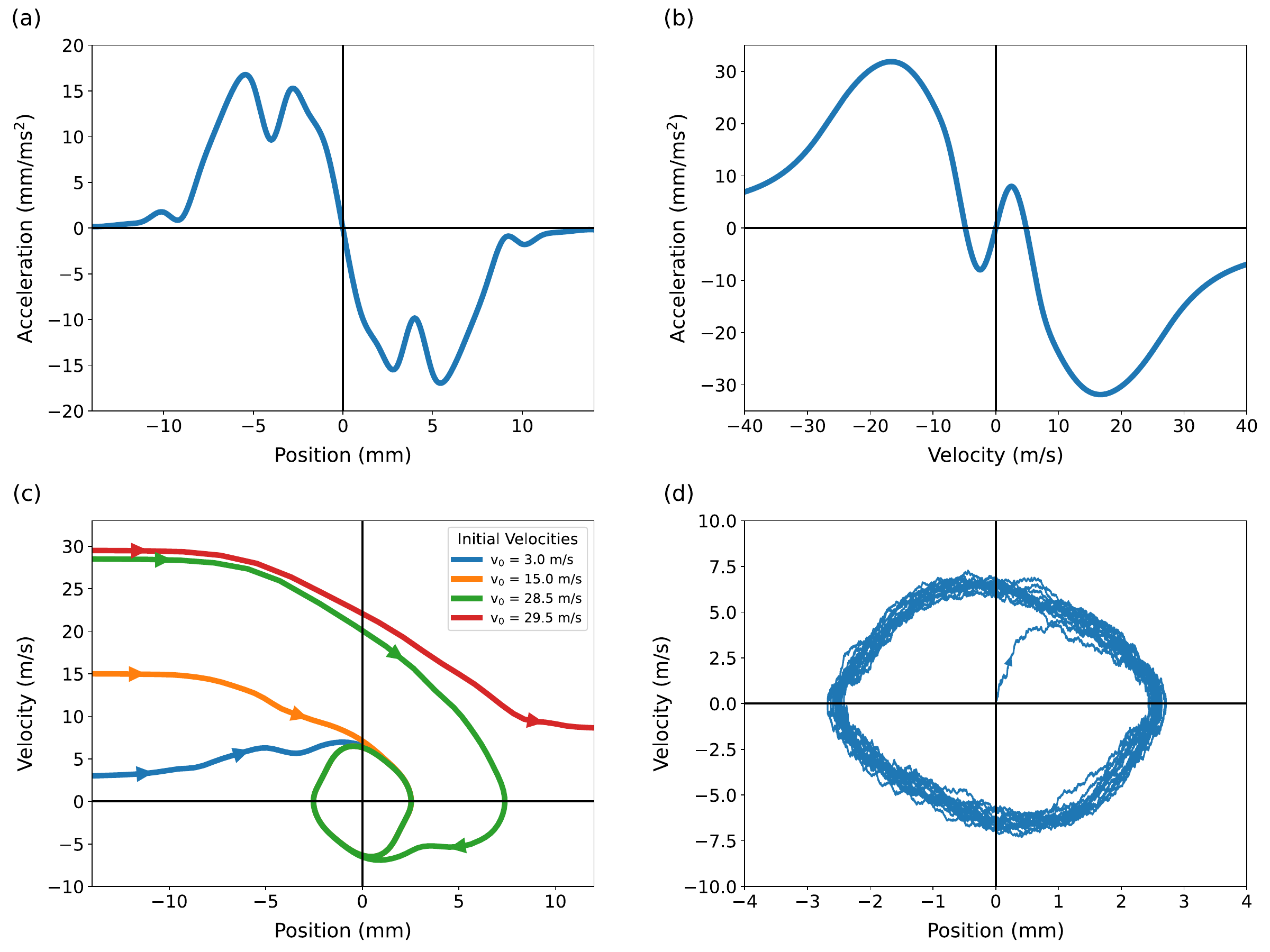}%
    \caption{Analysis of the simulated Sn capture MOT. (a) Acceleration $a$ vs displacement $r$. The curve shows positional restoring behavior to the center of the trap. (b) Acceleration $a$ vs velocity $v$. The curve shows strong Doppler cooling leading to velocity damping for high velocities, as well as strong sub-Doppler heating at low velocities. (c) Phase space trajectory of Sn atoms entering the MOT region from the source beam. The arrowheads indicate the direction of time. Capture is possible for velocities up to $v_c \sim\! 28.5$ m/s. (d) In-MOT trajectory of Sn atoms in the capture MOT, accounting for spontaneous photon emission due to optical cycling. The arrowhead indicates the direction of time. By computing the rms velocity and displacement averaged over the last 10 ms of the trajectory, we find the MOT temperature $T \sim\! 225$ mK and cloud size $\sigma \sim\! 1.9$ mm.}
    \label{fig: red MOT supplemental}
\end{figure*}

Numerical simulations of all Sn MOT stages used a geometry where the $z$-axis is defined by the direction of the symmetry axis of the quadrupole magnetic field for the MOT. MOT laser beams in the $xy$-plane (which define the $x$ and $y$ axes) are oriented 45 degrees with respect to the atomic beam, which defines the axis $\hat{r} = (\hat{x} + \hat{y})/\sqrt{2}$. We use the same methodology as in Ref.~\cite{lde2023} to analyze the behavior of the capture MOT, i.e. we only consider motion $\mathbf{v} \parallel \hat{r}$ and $\mathbf{d} \parallel \hat{r}$, where $\mathbf{v}$ and $\mathbf{d}$ denote 3-D velocity and displacement of atoms in the MOT. After solving the optical Bloch equations to obtain $a(r,v)$, we take one-dimensional cuts to elucidate the behavior of $a(r)$ and $a(v)$. These curves are defined by:
\begin{equation}
    a(r) = \frac{1}{2v_{ \text{max}}}\int_{-v_{ \text{max}}}^{v_{\text{max}}} a(r,v)\,dv,
\end{equation}
and:
\begin{equation}
    a(v) = \frac{1}{2r_\text{max}}\int_{-r_\text{max}}^{r_\text{max}} a(r,v)\,dr.
\end{equation}
Here, $v_{\text{max}}$ and $r_\text{max}$ are the maximum velocity and displacement of atoms in the capture MOT, which we find to be $v_\text{max}\approx 8$ m/s and $r_\text{max} \approx 3$ mm. Plots of these cuts are shown in Fig.~\ref{fig: red MOT supplemental}(a) and (b), indicating the trapping and cooling behavior of the capture MOT stage. The sign reversal in the $a(v)$ curve at low velocities is a hallmark of sub-Doppler heating, a signature behavior of Type-II MOTs.

Fig.~\ref{fig: red MOT supplemental}(c) shows how the capture velocity $v_c$ is determined, as described in the main text. We find that for velocities up to $v_c \sim 28.5$ m/s, atoms entering the MOT are slow enough to be captured. Captured atoms undergo a periodic, circular-like trajectory in phase-space, centered about the origin. This is indicative of sub-Doppler heating while trapped, due to the Type-II optical cycling transition. 

Fig.~\ref{fig: red MOT supplemental}(d) shows an example phase-space plot illustrative of how the temperature and size of the simulated capture MOT are determined. An atom is initialized at the phase-space origin and its trajectory propagated using the calculated acceleration $a(r,v)$. To account for random photon kicks from spontaneous emission during optical cycling, we give the atom a momentum kick in a random direction during each scattering event, which occurs with a rate $\sim P_{\rm exc}\Gamma$. The in-MOT trajectory with photon scattering is propagated for 50 ms to allow all observables to reach equilibrium. The temperature and size for each trajectory are calculated by computing the rms velocity and displacement in the last 10 ms of evolution. We average these values over 10 independent, stochastic trajectories to obtain the reported values for temperature $T$ and cloud size $\sigma$.

Compressed MOT simulations were analyzed using the same protocol as for the capture MOT, since the two stages are identical except for the reduced laser power, reduced laser detuning, and increased $B$-field gradient of the compression stage compared to the capture stage. The main difference in analysis is that a finer grid for interpolation of $a(r,v)$ values is used in the compression stage, to account for the fact that the MOT temperature and size reduces during this sequence. Conveyor belt blue MOT simulations were also analyzed similarly, with emphasis placed on the rms positional compression of the MOT cloud during this stage.

\section{Further Details on Tin White Light Slowing Simulations}\label{sec:AppxWLSSimDetails}
Numerical simulations of white light slowing (WLS) of Sn used a geometry where the atomic beam defines the $z$-axis. The slowing laser polarization is linear and defines the $x$-axis. Sn atoms exit the CBGB source with longitudinal velocity $v_z$ and transverse velocity $v_\perp$ drawn from independent Gaussian distributions, as described in the main text. Their transverse displacement $r$ is drawn from a probability distribution which is the convolution of a Gaussian of FWHM $\approx 4$ mm and a 2-D uniform disk of radius $R = 1.5$ mm centered on the $z$-axis. Sn atoms are initialized at $z_Q = 7.5$ mm, which defines the end of the ``zone of freezing" where re-thermalizing collisions of Sn atoms with He buffer gas cease. 

We consider realistic experimental parameters in our simulations of white light slowing. Spectral broadening at ultraviolet laser wavelengths can be achieved using resonant crystal electro-optic modulators (EOMs), whose modulation depths typically cannot exceed $\beta \sim 2$ rad. For sufficient laser broadening, we use a dual EOM configuration where the slowing laser passes through two EOMs in succession ($\beta_1 = \beta_2 = 1.5$ rad, $\Omega_1 = 1.5\Gamma$, $\Omega_2 = 3.4\Gamma$). The broadened laser spectrum can be analytically calculated using the Jacobi-Anger expansion. The time dependence of the laser electromagnetic field is given by:
\begin{widetext}
    \begin{multline}
    e^{i\omega t + i\beta_1\sin(\Omega_1 t) + i\beta_2\sin(\Omega_2 t)} = e^{i\omega t}\left(J_0(\beta_1) + \sum_{k=1}^\infty J_k(\beta_1) e^{ik\Omega_1 t} + \sum_{k=1}^\infty (-1)^k J_k(\beta_1) e^{-ik\Omega_1 t} \right) \times \\ \left(J_0(\beta_2) + \sum_{k=1}^\infty J_k(\beta_2) e^{ik\Omega_2 t} + \sum_{k=1}^\infty (-1)^k J_k(\beta_2) e^{-ik\Omega_2 t} \right),
\end{multline}
\end{widetext}
where $J_k(\beta)$ is the $k$th order Bessel function of the first kind. This corresponds to having spectral power in the original carrier $\omega$ as well as at frequency sidebands $\omega \pm k\Omega_1 \pm m\Omega_2$ where $k,m$ are integers. The powers in each spectral component are determined by the values of $J_k(\beta_1)$ and $J_m(\beta_2)$. 

\begin{figure}[h!]
    \includegraphics[width=\linewidth]{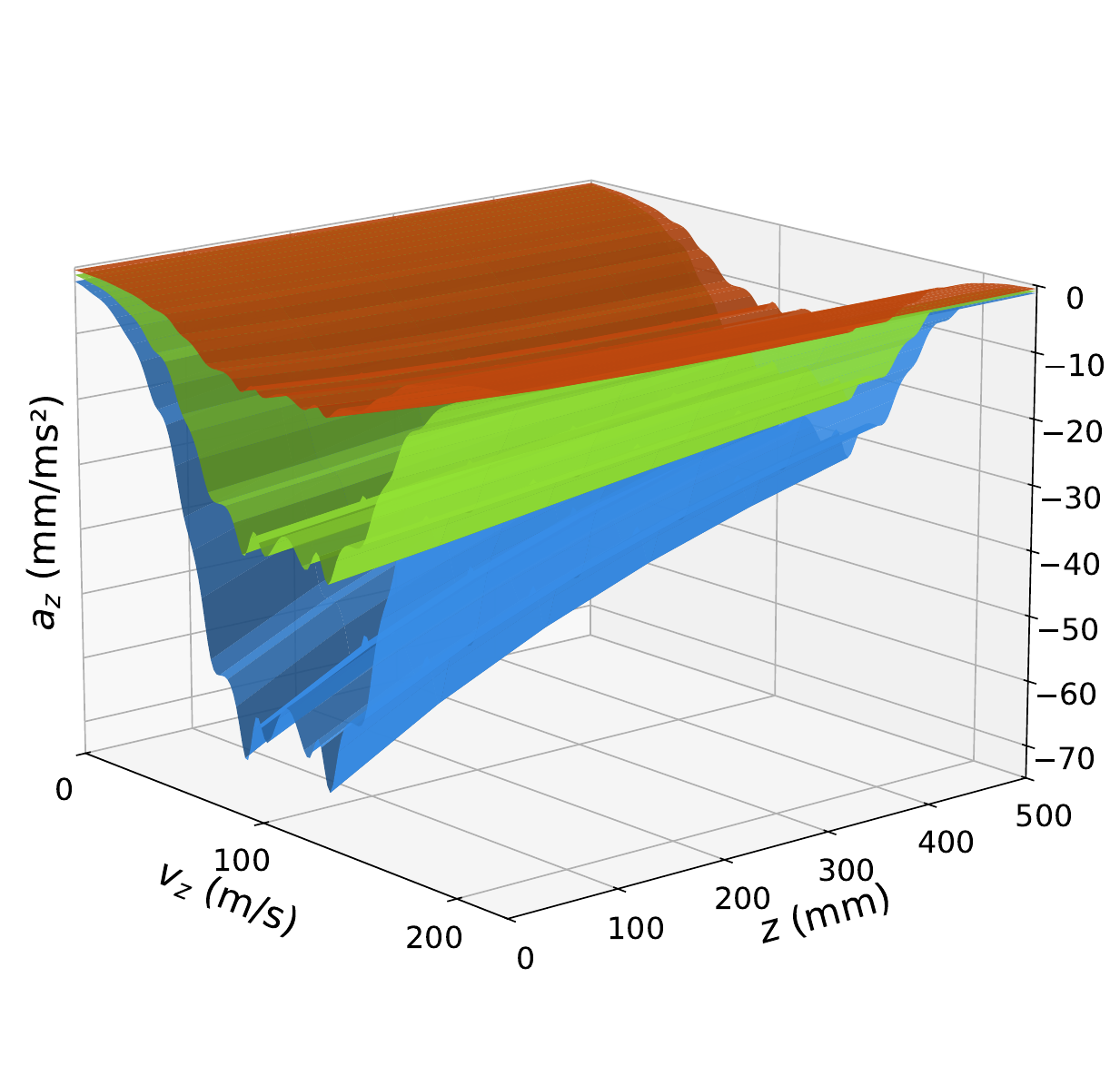}%
    \caption{Three-dimensional plot of the longitudinal deceleration $a_z(v_z, z)$ imparted on Sn atoms due to white light slowing. Profiles for three different transverse displacements from the atomic beam axis $\hat{z}$ are shown: $r=0$ mm (blue), $r=3$ mm (green), and $r=6$ mm (brown). These show the decrease in slowing force as $r$ increases. The longitudinal intensity gradient due to the converging slowing beam is apparent as the peak value of $a_z$ gets increasingly negative for decreasing $z$.}
    \label{fig:WLSAccelCurve}
\end{figure}

After applying this broadened laser spectrum in the WLS simulations, we solve the OBEs and obtain the longitudinal deceleration $a_z(v_z, r)$. We incorporate a converging slowing beam with half-angle $\theta' = 4.5$ mrad by solving the OBEs for $a_z(v_z, r)$ as a function of longitudinally-varying laser intensity $s(z)$. Additionally, we take the transverse deceleration due to the converging slowing beam to be $a_\perp = a_z \tan \theta'$. Interpolation across $z$ values from $z=0$ mm to $z=500$ mm enables us to obtain deceleration curves $a_z(v_z, r, z)$, which are shown in Fig.~\ref{fig:WLSAccelCurve}. The calculated $a_z(v_z, r, z)$ and $a_\perp(v_z, r, z)$ are used to solve the differential equations for the trajectory of Sn atoms as they are slowed from source to MOT region. 

\bibliography{bibtexMasterFile}
\bibliographystyle{apsrev4-2}

\end{document}